\RequirePackage{fix-cm}
\documentclass[smallextended]{svjour3}       
\smartqed  
\usepackage{graphicx}
\usepackage[T1]{fontenc}
\usepackage[latin9]{inputenc}
\usepackage{color}
\usepackage{float}
\usepackage{amsbsy}
\usepackage{amstext}
\usepackage{graphicx}
\usepackage{esint}
\usepackage{times}

\usepackage{bm}
\usepackage{array}
\usepackage{epstopdf}

\newcommand{\HD}{\mathop{H_\textrm{D}}}
\newcommand{\HC}{\mathop{H_\textrm{C}}}
\newcommand{\HDC}{\mathop{H_\textrm{DC}}}
\newcommand{\VDC}{\mathop{V_\textrm{DC}}}

\newcommand{\epsd}{\mathop{\epsilon_{\textrm{d}}}}

\newcommand{\zc}{\mathop{\bar{z}_\textrm{c}}}

\newcommand{\epspm}{\bar{\epsilon}_\pm}
\newcommand{\epsm}{\bar{\epsilon}_-}
\newcommand{\epsp}{\bar{\epsilon}_+}

\newcommand{\bepsd}{\bar{\epsilon}_\textrm{d}}

\newcommand{\im}{\text{Im}}

\begin{document}

\title{Analysis technique for exceptional points in open quantum systems and QPT analogy for the appearance of irreversibility}


\author{
Savannah Garmon         \and
      Ingrid Rotter      \and
        Naomichi Hatano        \and
       Dvira Segal
}


\institute{
		S. Garmon   \and D. Segal \at
		Chemical Physics Theory Group, Department of Chemistry and Center
		for Quantum Information and Quantum Control, University of Toronto,
		80 St. George Street, Toronto, Ontario, Canada M5S 3H6
           \and
           I. Rotter \at
		Max-Planck-Institut f\"ur Physik Komplexer Systeme,
		D-01187 Dresden, Germany
	\and
	N. Hatano  \at
		Institute of Industrial Science,
		University of Tokyo, Komaba 4-6-1,
		Meguro, Tokyo 153-8505, Japan
	\and
	S. Garmon
	    \email{sgarmon@chem.utoronto.ca}
}

\date{Received: date / Accepted: date}

\maketitle

\global\long\def\ket#1{\left| #1\right\rangle }

\global\long\def\bra#1{\left\langle #1 \right|}

\global\long\def\kket#1{\left\Vert #1\right\rangle }

\global\long\def\bbra#1{\left\langle #1\right\Vert }

\global\long\def\braket#1#2{\left\langle #1\right. \left| #2 \right\rangle }

\global\long\def\bbrakket#1#2{\left\langle #1\right. \left\Vert #2\right\rangle }

\global\long\def\av#1{\left\langle #1 \right\rangle }

\global\long\def\tr{\text{Tr}}

\global\long\def\im{\text{Im}}

\global\long\def\re{\text{Re}}

\global\long\def\sign{\text{sgn}}

\global\long\def\abs#1{\left|#1\right|}

\begin{abstract}
We propose an analysis technique for the exceptional points (EPs) occurring in the discrete spectrum of open quantum systems (OQS), using a semi-infinite chain coupled to an endpoint impurity as a prototype.  We outline our method to locate the EPs in OQS, further obtaining an eigenvalue expansion in the vicinity of the EPs that gives rise to characteristic exponents.
We also report the precise number of EPs occurring in an OQS with a continuum described by a quadratic dispersion curve.
In particular, the number of EPs occurring in a bare discrete Hamiltonian of dimension
$n_\textrm{D}$ is given by 
$n_\textrm{D} (n_\textrm{D} - 1)$; if this discrete Hamiltonian is then coupled to continuum (or continua) to form an OQS, the interaction with the continuum generally produces an enlarged discrete solution space that includes a greater number of EPs,
specifically 
$2^{n_\textrm{C}} (n_\textrm{C} + n_\textrm{D} )  
[ 2^{n_\textrm{C}} (n_\textrm{C} + n_\textrm{D} )   - 1] $,
in which $n_\textrm{C}$ is the number of (non-degenerate) continua to which the discrete sector is attached.
Finally, we offer a heuristic quantum phase transition analogy for the emergence of the resonance (giving rise to irreversibility via exponential decay) in which the decay width plays the role of the order parameter; the associated critical exponent is then determined by the above eigenvalue expansion.
\end{abstract}


\section{Introduction}\label{sec:intro}

Exceptional points (EPs) represent defective points occurring in the parameter space of some finite-dimensional Hamiltonian $\HD$ at which two (or possibly more) of the system eigenvalues coalesce while, simultaneously, the usual self-adjoint property of the Hamiltonian fails to hold \cite{Heiss99}.
EPs serve to characterize the qualitative aspects of the spectrum in their vicinity and have been studied in a wide array of physical contexts, including experimental studies of the field modes in a microwave cavity \cite{EPexpt1a,EPexpt1b} as well as a chaotic optical microcavity \cite{EPexpt2}.  
In the context of scattering theory an EP at which two eigenvalues coalesce appears as a double pole of the S-matrix; for examples see the analysis of nuclear spectra, laser-induced continuum structures in atoms and the transmission through quantum dots that is reported in the review
 \cite{Rotter_review}.
Further theoretical studies of exceptional points in various physical contexts include superconducting wires \cite{RSM07}, electron scattering on helium \cite{MF80}, photodissociation of $H_2^+$ \cite{LASM09}, the hydrogen spectrum under crossed electric and magnetic fields \cite{CMW07}, the scattering of a beam of particles in a double barrier potential \cite{HJM06}, and nuclear spectra studies relying on random matrix theory \cite{ZVW83}.  
A quantum phase transition (QPT) interpretation has also been offered based on a Coulomb interaction analogy for the EPs \cite{CHM07}.

In physical situations the finite Hamiltonian $\HD$ is usually embedded in a much larger system (i.e. open quantum system (OQS)) such that the discrete spectrum of $\HD$ is perturbed by the continuous degrees of freedom associated with the larger system.  The continuum might represent a measuring device, such as the leads used to probe the cavity modes in Refs. \cite{EPexpt1a,EPexpt1b}.
Note, however, that the EPs in this case are not merely shifted from those of the discrete space by itself; in fact, the interaction with the continuum in general gives rise to a greater number of EPs inhabiting an effective finite-dimensional Hamiltonian with dimension larger than $\HD$ alone (we demonstrate this subtle point for our prototype model below).

Here we propose a method for the discovery and spectral analysis of EPs in OQS.  This analysis incorporates a certain eigenvalue expansion that generically occurs in the vicinity of any EP \cite{Kato,RSM07,MF80}.
Further, we offer our own QPT analogy that relates to the emergence of the resonant state in an open system and hence acts as a 
mechanism for spontaneous time-symmetry breaking in nature,
with critical exponents given by this eigenvalue expansion.  
While we will demonstrate this process for a specific example, we emphasize that a similar effect occurs in many more general OQS models.

The Hamiltonian we have in mind may be generically written as $H = \HD + \HC + \HDC$; here $\HC$ gives the continuum (open) component of the system coupled to a discrete level or series of levels 
$\HD$ through the coupling Hamiltonian $\HDC$.  We will focus on the case 
$\HD = \epsd d^\dagger d$, in which the discrete component is written in terms of a single level coupled to the continuum via $\HDC = g \VDC$ with $g$ a dimensionless coupling constant.  
However, we make this specification with the understanding that
this single level may simply be one particular discrete level, with others absorbed into $\HC$, and the general ideas from our analysis below still applies in the case of multiple levels \cite{Heiss99}.  We assume that no many-body interactions are present.

This Hamiltonian (or close generalizations) may be applied in the description of a wide range of physical phenomena, for example the interaction of a two-level oscillator with the field modes in a waveguide \cite{PTG05} or the effect of a quantum dot impurity on quantum wires \cite{Longhi09,GNHP09} and
has been studied extensively 
\cite{Rotter_review,BRS06,JMR99,LonghiPRL06,Longhi07,TGOP07}.
We comment on the expected general applicability of our method below.

Defining the Green's function $G(z) \equiv \left( z - H \right)^{-1}$ we obtain the discrete spectrum
for $H$ as the poles of 
$\bra d G(z) \ket d$, where $\ket d$ is the eigenvector of $\HD$ with the eigenvalue $\epsd$.  
For the simplest cases, this spectrum appears as the
solutions of a dispersion equation in the form
\begin{equation}
z - \epsd = \Sigma(z).
\label{disc.disp}
\end{equation}
with the self-energy function $\Sigma (z) \propto g^2$.
The open system allows for multiple solutions that may include complex solutions (i.e. resonances)  
\cite{HSNP08}.

We outline our analysis technique in this paper for the case that the dispersion equation is given by Eq. (\ref{disc.disp}).  While our method is oriented towards analytically solvable models, it nevertheless should prove useful in some numerical situations assuming that we may at least explicitly write the self-energy function $\Sigma (z)$ in integral form.  Further, our method should in principle be portable to other OQS such as the scattering problem in Ref. \cite{HJM06}.  For systems involving a more complicated structure for $\HD$ with multiple bound states, one generally obtains a matrix equation 
\cite{SHO11} equivalent of Eq. (\ref{disc.disp}), in which case it should again be possible to generalize the methods described below.  However, for systems involving significantly more complex interactions between the discrete sector and the fields, such as Ref. \cite{CMW07}, this may be difficult to implement in practice, as well for the case involving multi-particle interactions.

Let $x$ generically represent any two Hamiltonian parameters, such as $x = \{ g, \epsd \}$ above.  
An EP is a point $\bar{x}$ in parameter space
at which two eigenvalues $z_i (x)$ and $z_j (x)$ (though possibly more) coalesce as $z_i (\bar{x}) = z_j (\bar{x}) \equiv \zc$, while their eigenvectors also collapse into one another.
Note that we have defined $\zc$ as the coalescence point in the complex energy plane.  The coalescence occurs because the two eigenvalues share a common branch cut in the complex parameter space.  The branch cut is described by the system discriminant $D(x)$, which is obtained in the case of a closed system from the determinant of the finite Hamiltonian; in the case of an OQS we may obtain $D(x)$ 
as the determinant of a finite-dimensional effective Hamiltonian 
$H_\textrm{eff}$ that describes the spectrum in the discrete subspace 
($H_\textrm{eff}$ may be obtained by applying the Feshbach projection operator technique \cite{Rotter_review} or by other methods \cite{HSNP08,Hatano_H_eff} as we demonstrate in App. \ref{app:H.eff}).  Adiabatic encirclement of the defective point $\bar{x}$ in complex parameter space will exchange the two eigenvalues \cite{Heiss99,CMW07,Kato}.

In a more general context, following Kato's formalism \cite{Kato} we may group the system eigenvalues in the vicinity of a given EP $\bar{x}$ into \emph{cycles} as
\begin{equation}
\{ z_1 (x), \dots, z_{p-1} (x), z_p (x) \}, \; \;
\{ z_{p+1} (x), \dots, z_{p+q} (x) \},\dots
\label{cycles}
\end{equation}
Upon a single adiabatic revolution of the system parameters about $\bar{x}$ the eigenvalues will be mapped into one another according to
\begin{equation}
\{ z_1 (x), \dots, z_{p-1} (x), z_p (x) \} \rightarrow \{ z_2 (x), \dots, z_p (x), z_1 (x) \}
\label{cycles.iter}
\end{equation}
and so forth.  Hence we must complete $p$ revolutions in order for this cycle to return to its original configuration, where $p$ is the \emph{period} (the number of elements in a given cycle).  
Note that we have $p = 2$, the simplest case, for our prototype model starting from Sec. \ref{sec:proto}; however we maintain that $p$ may take on arbitrary values $p \ge 2$ for the time being.
Further, we call the energy value $\zc$ 
at which the eigenvalues coalesce the \emph{center} of the cycle. 

We offer that the elements of the cycle may be expanded in the vicinity of the EP in a fractional power series in the form
\begin{equation}
z_h (x) = \zc + \alpha_{1,h} \left( f_j (x) \right)^{1/p} 
				+ \alpha_{2,h} \left( f_j (x) \right)^{2/p} + \dots
\label{kato.puiseux}
\end{equation}
Here $\alpha_{l, h} = \beta_l \exp(2 \pi i h / p)$ with $\beta_l$ an unknown coefficient to be determined below and $h= 0, 1, \dots p-1$.
The functions $f_j (x)$ are polynomials obtained by factoring the discriminant $D(x)$ such that EPs associated with a shared branch cut have been placed in a common factor (an example appears below).
Our expression Eq. (\ref{kato.puiseux})
is a mild generalization of that appearing in Ref. \cite{Kato}, in which only linear polynomials 
$f_j (x) =  x - \bar{x} $ are considered.
Note that in the case of an ordinary Taylor expansion the distinction would be largely irrelevant, as in that case the expression incorporating higher order polynomials could just as well be re-grouped to give the more familiar form written in terms of linear factors.  However, the fractional powers appearing in Eq. (\ref{kato.puiseux}) make such a re-grouping impossible in the present case; hence here it is best to work with the most general expression possible.

Having introduced our essential framework for the EPs, we proceed as follows.  In Sec. \ref{sec:anal.tech} we briefly present our technique for locating and analyzing EPs in OQS.  We turn in Sec. \ref{sec:proto} to the specific case of a semi-infinite tight-binding chain with an endpoint impurity in order to provide an example where the EP analysis may be performed in a simple analytic context with closed-form expressions.  
In Sec. \ref{sec:spectrum} we present a detailed look at the eigenvalue spectrum for the case $g < 1/\sqrt{2}$, in which a resonance is present in our prototype model for certain values of $\epsd$.  We further obtain the fractional power series Eq. (\ref{kato.puiseux}) for our prototype for values of $\epsd$ in the vicinity of the point in parameter space at which the resonance appears.  
We also take the opportunity to comment on the number of EPs occurring in OQS, dependent on the form of the dispersion in the leads.
In Sec. \ref{sec:qpt} we present our QPT analogy for the appearance of the resonance and the consequent breaking of time-reversal symmetry.
We summarize our work and make final conclusions in Sec. \ref{sec:conc}.

We also demonstrate how to obtain an isospectral effective Hamiltonian for our OQS in App. \ref{app:H.eff}, and in App. \ref{app:H.eff.n.solns} we comment on the number of solutions appearing in an OQS described by a given effective Hamiltonian.
The discussion around the effective Hamiltonian elucidates the relationship between the typical OQS we have in mind and the $\mathcal{PT}$-symmetric Hamiltonians that have been the subject of numerous studies in the last decade or so.


\section{Analysis technique for EPs in OQS}\label{sec:anal.tech}

We now briefly outline our general method for finding the EPs in OQS as well as finding the analytic form of the expansion Eq. (\ref{kato.puiseux}).
We emphasize that this technique should remain useful as a numerical tool in the case of systems significantly more complex than our prototype model introduced in Sec. \ref{sec:spectrum}.
We comment on some other methods for locating the EPs at the end of this section.

For our approach, we take the derivative of Eq. (\ref{disc.disp}) with respect to the impurity energy $\epsd$ and re-arrange to obtain
\begin{equation}
\frac{\textrm{d} \Sigma (z_0)}{\textrm{d} z_0}	=
	1 - \frac{1}{\partial z_0 / \partial \epsilon_\textrm{d}} .
\label{disc.disp.diff}
\end{equation}
As the eigenvalue derivative diverges at the EP \cite{MF80} according to $\partial z_i / \partial \epsd |_{\epsd = \bar{\epsilon}_\textrm{d}} \rightarrow \infty$ (which can be seen immediately from Eq. (\ref{kato.puiseux}) since $p >1$) we may evaluate Eq. (\ref{disc.disp.diff}) at the EP as
\begin{equation}
\left.  \frac{\textrm{d} \Sigma (z_0)}{\textrm{d} z_0} \right|_{z_0 = \zc}
	= 1 ,
\label{self.energy.cond.center}
\end{equation}
where $\zc$ is the center associated with the EP $\bar{\epsilon}_\textrm{d}$.
This gives us a condition to determine the value of the center $\zc$ in terms of 
the impurity-continuum coupling parameter $g$ 
\footnote{Note that Eq. (\ref{self.energy.cond.center}) is the equivalent of Eq. (16) in Ref. \cite{HJM06}; hence the technique described here below Eq. (\ref{self.energy.cond.center}) could be carried over to the scattering problem as well. }.
We determine the location of the EP in the parameter space $\left( \epsd, g \right)$ 
by plugging $\zc$ into the dispersion 
Eq. (\ref{disc.disp}) and solving for $\bepsd$ for a fixed value of $g$ (or vice versa $\bar{g}$ for 
fixed $\epsd$); this result naturally occurs as one of the discriminant polynomial factors $f_j (\epsd, g)$
in Eq. (\ref{kato.puiseux}).
While finding the EP, note that a sign ambiguity will arise associated with the branch cut; this can be overcome by performing a consistency check on the value of $\zc$ obtained in 
Eq. (\ref{self.energy.cond.center}).

Having obtained $\zc$ along with the relevant EP,
we further obtain the higher order terms in 
Eq. (\ref{kato.puiseux}) to arbitrary precision by expanding Eq. (\ref{disc.disp.diff}) in powers of $\left( f_j (\epsd, g) \right)^{1/p}$ 
to obtain conditions on the coefficients $\alpha_{l, h}$ in terms of the system parameters.  The remaining unknown is the period $p$, which is obtained by integrating the logarithm of the Green's function $G(z)$ around a closed contour $\mathcal C_{\zc}$ in (energy) $z$-space about 
$\zc$ as
\begin{equation}
p
	=   \frac{1}{2 \pi i} \oint_{\mathcal C_{\zc}}
			\frac{\partial}{\partial z}  \bra d  \log G(z) \ket d  dz .
\label{p.fcn}
\end{equation}
Note that once the location of the EPs are determined, the value of $p$ could be found numerically in a straight-forward manner using this equation.

Clearly the above method for locating the EPs is only useful in the case that we can write some type of closed form expression for $\Sigma (z)$.  In the case that this is not possible, several purely numerical approaches have been proposed in the literature \cite{JKM10,LM10,UL10}.  However, the numerical method presented in Ref. \cite{JKM10} requires a finite matrix  as input, while an OQS necessarily contains a continuum element.  This problem may be overcome by introducing an isospectral finite-dimensional effective Hamiltonian $H_\textrm{eff}$, as in App. \ref{app:H.eff}.  This introduces yet another issue: the method in Ref. \cite{JKM10} assumes the relevant eigenvalue equation takes the form $H \psi_z = z \psi_z$ while the effective Hamiltonian method results in a more generalized eigenvalue equation of the form $H_\textrm{eff} (z) \psi_z = z \psi_z$, as demonstrated in App. \ref{app:H.eff}.  Hence one would need to check carefully if this method can be appropriately generalized.


\section{Prototype model: application to a semi-infinite chain with end-point impurity}
\label{sec:proto}

\begin{figure}
 \includegraphics[width=0.9\textwidth]{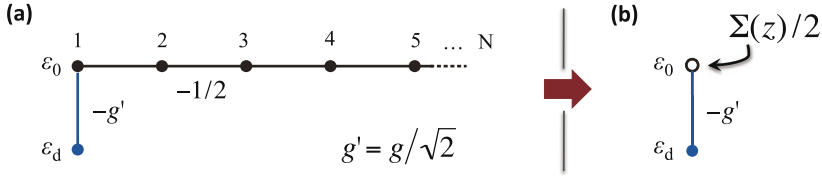}
 \caption{
(a) Prototype model geometry: semi-infinite chain coupled with a discrete endpoint impurity; the nearest-neighbor coupling is given by $-1/2$ and the chemical potential of the sites along the chain is 
$\epsilon_0 = 0$.  The impurity has chemical potential $\epsilon_\textrm{d}$ and is attached to the endpoint site $1$ with coupling $g' = g / \sqrt{2}$.  (b) Diagrammatic representation of the finite-dimensional effective Hamiltonian $H_\textrm{eff}$ (z) derived in App. \ref{app:H.eff}.  The effect of the chain on the impurity site is captured here by the self-energy function $\Sigma(z)$.
}
 \label{fig:semi.geo}
\end{figure}

We now apply the techniques above to a semi-infinite chain with an end-point impurity as an illustration of the emergence of a resonant state, which occurs at a real-valued exceptional point.  Further on we will offer our QPT analogy for this process, with the characteristic exponent for the order parameter (decay width) read off from Eq. (\ref{kato.puiseux}).
Our model Hamiltonian (see Fig. \ref{fig:semi.geo}(a)) takes the form
\begin{equation}
H = \epsilon_\textrm{d} d^\dagger d 
	- \frac{1}{2} \sum_{i=1}^N \left( c_n^\dagger c_{n+1} + c_{n+1}^\dagger c_n \right)
	- \frac{g}{\sqrt{2}} \left( c_{1}^\dagger d + d^\dagger c_1 \right) ,
\label{ham.semi.site}
\end{equation}
in which the coefficient of the second term sets the energy scale for this paper.
This model is exactly solvable and has been the subject of numerous studies \cite{Longhi09,Pastawski_footnote,LonghiPRL06,Garmon_non-Mark}.
The site-to-site hopping in the second term of Eq. (\ref{ham.semi.site}) is easily diagonalized by introducing a Fourier transform
\begin{equation}
c_k^\dagger = \sqrt{\frac{2}{\pi}} \sum_{i=1}^N \sin nk \ c_n^\dagger
	,
\label{semi.fourier}
\end{equation}
which reveals the dispersion $\epsilon_k = - \cos k$,
covering $k \in \left[0, \pi \right]$ in the continuum limit $N \rightarrow \infty$.
The third term gives the impurity coupling to the end-point of the chain with strength $- g/\sqrt{2}$.
We obtain the discrete spectrum for the fully diagonalized model from the poles of 
$\bra d G(z) \ket d = \left( z - \epsilon_\textrm{d} - \Sigma (z) \right)^{-1}$.
Since our model is purely quadratic, this expression may be evaluated by a re-summation following a straight-forward operator expansion in terms of the coupling.  Performing this evaluation
yields the equation for the poles as
\begin{equation}
z - \epsilon_\textrm{d} = \Sigma (z) = g^2 \left( z - \sqrt{z^2 - 1} \right)
\label{semi.disp}
\end{equation}
(we present an alternative derivation of this result in App. \ref{app:H.eff}). 
We may square out the root in this expression to obtain a quadratic $P_\textrm{q}(z)$, then 
following Ref. \cite{Fuchs&Levin} we obtain the system discriminant
from the solutions of $P_\textrm{q}(z) = P_\textrm{q}^\prime (z)= 0$ as
\begin{equation}
D (\epsd ; g )
	= - 4 g^4 f_1 (g) f_2 (\epsd,g) ,
\label{semi.disc}
\end{equation}
in which we define the polynomial factors $f_1 (g) \equiv 1 - 2g^2$ and
$f_2 (\epsd, g) \equiv  \epsd^2 - (1 - 2g^2)$ mentioned above.
The polynomial $f_2  (\epsd, g)$ describes a branch cut in complex $\epsd$ space extending from 
$\epsm$ to $\epsp$ with 
\begin{equation}
\epspm \equiv \pm \sqrt{1 - 2g^2} ,
\label{semi.EPs}
\end{equation}
revealing $\epsd = \epspm$ as two EPs 
written in terms of fixed $g$.
We note immediately that for $g < 1/\sqrt{2}$ ($g > 1/\sqrt{2}$) 
these EPs are real (imaginary); for present purposes we will focus on the real case.
Meanwhile the polynomial $f_1 (g)$ reveals that 
$g = 1/\sqrt{2}$ is (coincidentally) some special point in its own right (this will be further considered in a separate publication).

Before proceeding with a closer inspection of the eigenvalues, we take advantage of an opportunity here to comment on the number of EPs appearing in certain kinds of OQS.  According to Ref. \cite{Kato}, the number of EPs $r$ appearing in the spectrum of a discrete Hamiltonian of dimension $n$ is given by $r = n (n -1)$.  However, the situation for OQS is a bit more subtle.  In the case of our prototype system, we immediately notice that the number of EPs is \emph{not} fixed by the dimensionality of the bare discrete sector 
$\HD$, according to which there would be zero EPs (after all, a single eigenvalue cannot coalesce into itself).  To elucidate the situation, it is useful to consider the effective Hamiltonian (see Fig. \ref{fig:semi.geo}(b))
\begin{equation}
H_\textrm{eff} (z) = 
	\left[ \begin{array}{cc}
		\epsd	& - \frac{g}{\sqrt{2}}	\\
		- \frac{g}{\sqrt{2}} &	\frac{1}{2} \Sigma (z) 
	\end{array}
	\right] ,
\label{semi.H.eff}
\end{equation}
which may be derived from the full Hamiltonian Eq. (\ref{ham.semi.site})
and captures many essential elements of the model; for example, the dispersion Eq. (\ref{semi.disp}) can be found immediately from 
$\det \left( H_\textrm{eff} (z) - z \right) = 0$ (see Ref. \cite{Hatano_H_eff}).  
The process of deriving this effective Hamtilonian 
Eq. (\ref{semi.H.eff}) from the full Hamiltonian for our prototype model is outlined in App. \ref{app:H.eff} and represented diagrammatically in the transition from Fig. \ref{fig:semi.geo}(a) to (b).

From here, one might naively expect that the number of EPs in OQS is now determined by the dimension $n_\textrm{eff}$ of $H_\textrm{eff}$ according to $r = n_\textrm{eff} \left( n_\textrm{eff} - 1 \right)$ and indeed this (by chance) gives the correct answer for our prototype model from $n_\textrm{eff} = 2$.
However, this is misleading; one must emphasize that $H_\textrm{eff}$ itself contains $z$-dependence, which results in a characteristic equation of the form $H_\textrm{eff} (z) \psi_z = z \psi_z$ that is usually of higher order than the dimension of $H_\textrm{eff}$ itself.  In fact, it can be shown that in the case of an effective Hamiltonian $H_\textrm{eff}$ of dimension $n$ with quadratic dispersion in the leads, the number of eigenvalues is given by $2n$  and that the present model is just a special case in which two of the solutions are trivial (see App. \ref{app:H.eff.n.solns} of the present text as well as Apps. D and H of \cite{SHO11}).

Indeed, observing the form of Eq. (\ref{semi.H.eff}) it is straight-forward to realize that the dimension of 
$H_\textrm{eff}$ increases by 1 for each discrete impurity we add to the system \emph{or} for each distinct chain (continuum) we add.  However, we must take into account that introducing a new chain also doubles the number of polynomial eigenvalue equations that must be solved, effectively doubling the number of solutions.  Hence the number of solutions in the general case is given by $2^{n_\textrm{C}} (n_\textrm{D} + n_\textrm{C})$ in which $n_\textrm{D}$ is the number of discrete impurity sites in $\HD$ and $n_\textrm{C}$ is the number of (non-degenerate) continuum
channels\footnote{For an example of a multi-channel model, consider the two-chain model in \cite{GNHP09}, in which it is straight-forward to derive the effective Hamiltonian $H_\textrm{eff}$ of dimension 3, giving rise to 12 solutions.}
in $\HC$.
Therefore, the number of EPs in OQS with quadratic dispersion is given by 
\begin{equation}
r = 2^{n_\textrm{C}} (n_\textrm{D} + n_\textrm{C}) [2^{n_\textrm{C}} (n_\textrm{D} + n_\textrm{C}) - 1]
.
\label{r.expression}
\end{equation}
For easy-to-check examples consider the systems in Refs. \cite{GNHP09,Longhi07,TGOP07}.

Returning now to our present evaluation, Eq. (\ref{semi.disp}) can be immediately solved for the two eigenvalues
\begin{equation}
z_\pm (\epsilon_\textrm{d}, g^2)
	= \epsilon_\textrm{d} \frac{1 - g^2}{1 - 2 g^2} 
		\pm g^2 \frac{ \sqrt{\epsilon_\textrm{d}^2 - \left(1-2g^2 \right)} }{1 - 2g^2}.
\label{semi.z.pm}
\end{equation}
We write the associated wave functions along the chain sites in the (unnormalized) form 
$\psi_\pm (x) \equiv \bra{\psi_\pm} x \rangle = e^{i k_\pm x} $ \cite{HSNP08} with
the effective $k$-value defined as $k_\pm \equiv \cos^{-1} (-z_\pm)$ and given explicitly as
\begin{equation}
k_\pm
	= i \log \left(  - \epsilon_\textrm{d} \pm \sqrt{\epsilon_\textrm{d}^2 - (1 - 2g^2)}	\right)  .
\label{semi.k.pm}
\end{equation}
Note that at the EPs we have $k_+ = k_-$ such that the two eigenfunctions collapse into one another.

Focusing on the EPs $\epsd = \epspm$, we briefly illustrate the method outlined for analyzing EPs in OQS by plugging $\Sigma (z)$ from Eq. (\ref{semi.disp}) into the condition Eq. (\ref{self.energy.cond.center}) to write
\begin{equation}
\Sigma ' (\bar{z}) = g^2 \left( 1 - \frac{\bar{z}}{q \sqrt{\bar{z}^2 - 1}} \right) = 1
\label{semi.disp.cond.center}
\end{equation}
as a condition to find the two values for the center, with $q=\pm$ a sign designation to be fixed below.  Equation (\ref{semi.disp.cond.center}) easily yields
\begin{equation}
\bar{z}_\textrm{c}^\pm = \pm \frac{1 - g^2}{\sqrt{1 - 2 g^2}}.
\label{semi.center.g}
\end{equation}
We perform a consistency check to find that the sign $q = +$ ($q = -$) before the root in 
Eq. (\ref{semi.disp.cond.center}) is associated with $\bar{z}_\textrm{c}^-$ ($\bar{z}_\textrm{c}^+$).
Making consistent use of this sign convention and plugging the value $z = \bar{z}_\textrm{c}^\pm$ into the original dispersion Eq. (\ref{semi.disp}) yields the condition for the EPs in parameter space as $\epsd = \epspm$, just as we found in Eq. (\ref{semi.EPs})
 from
$f_2 (\epsd, g) = 0$.  Finally we use Eq. (\ref{semi.center.g}) to write the center values in terms of the other system parameter $\epsd$ as $\bar{z}_\textrm{c}^\pm = \pm (1 + \epsd^2) / 2 \epsd$.   
From here we could continue the method outlined in the text following Eq. (\ref{self.energy.cond.center}) to obtain the eigenvalue expansion Eq. (\ref{kato.puiseux}). 
For a model with a more complicated self-energy function $\Sigma (z)$ this would probably be necessary but in the present case we may obtain the full expansion more easily.


\section{Prototype spectrum and eigenvalue expansion}
\label{sec:spectrum}

\begin{figure}
 \includegraphics[width=0.45\textwidth]{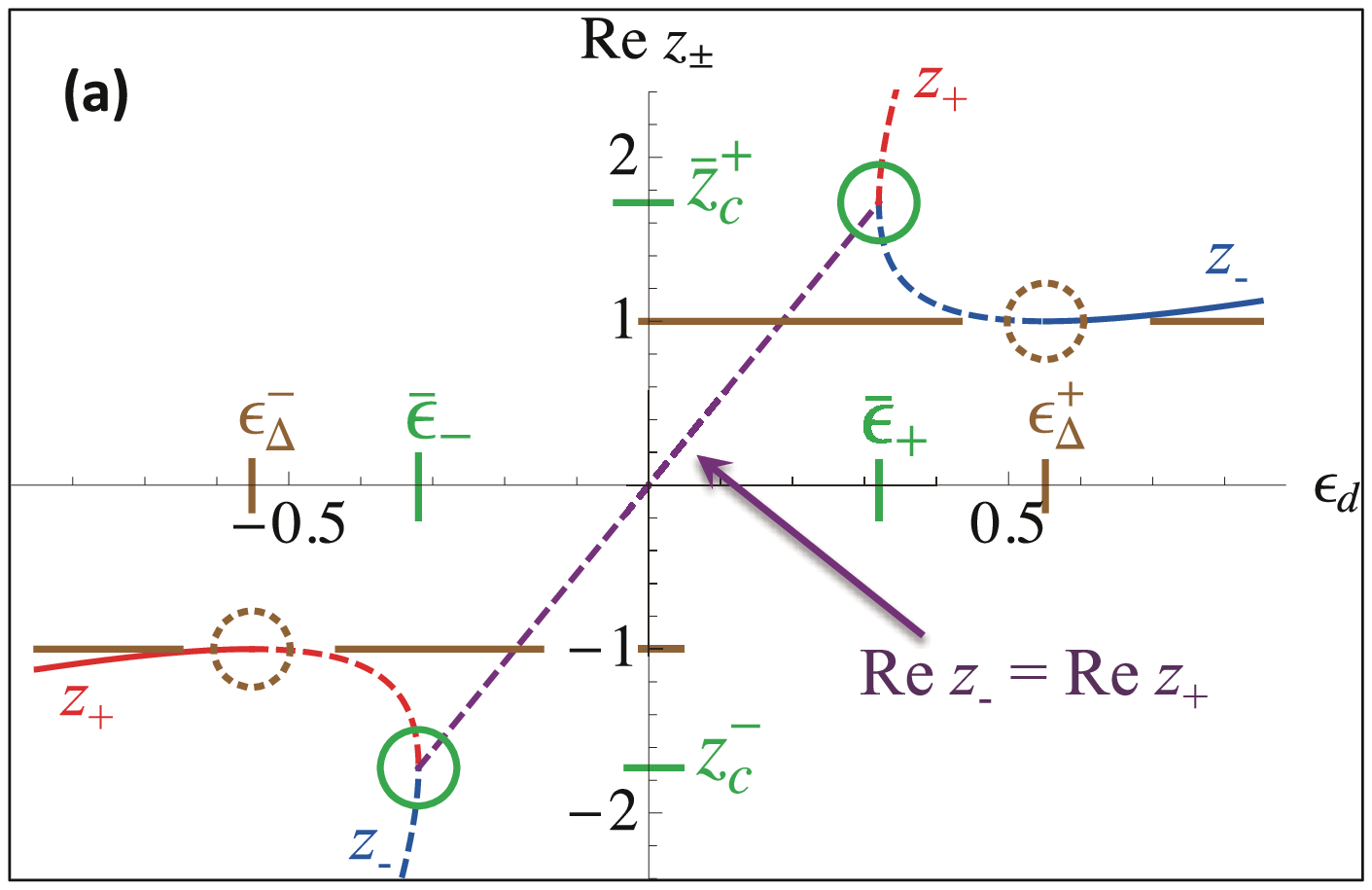}
 \includegraphics[width=0.45\textwidth]{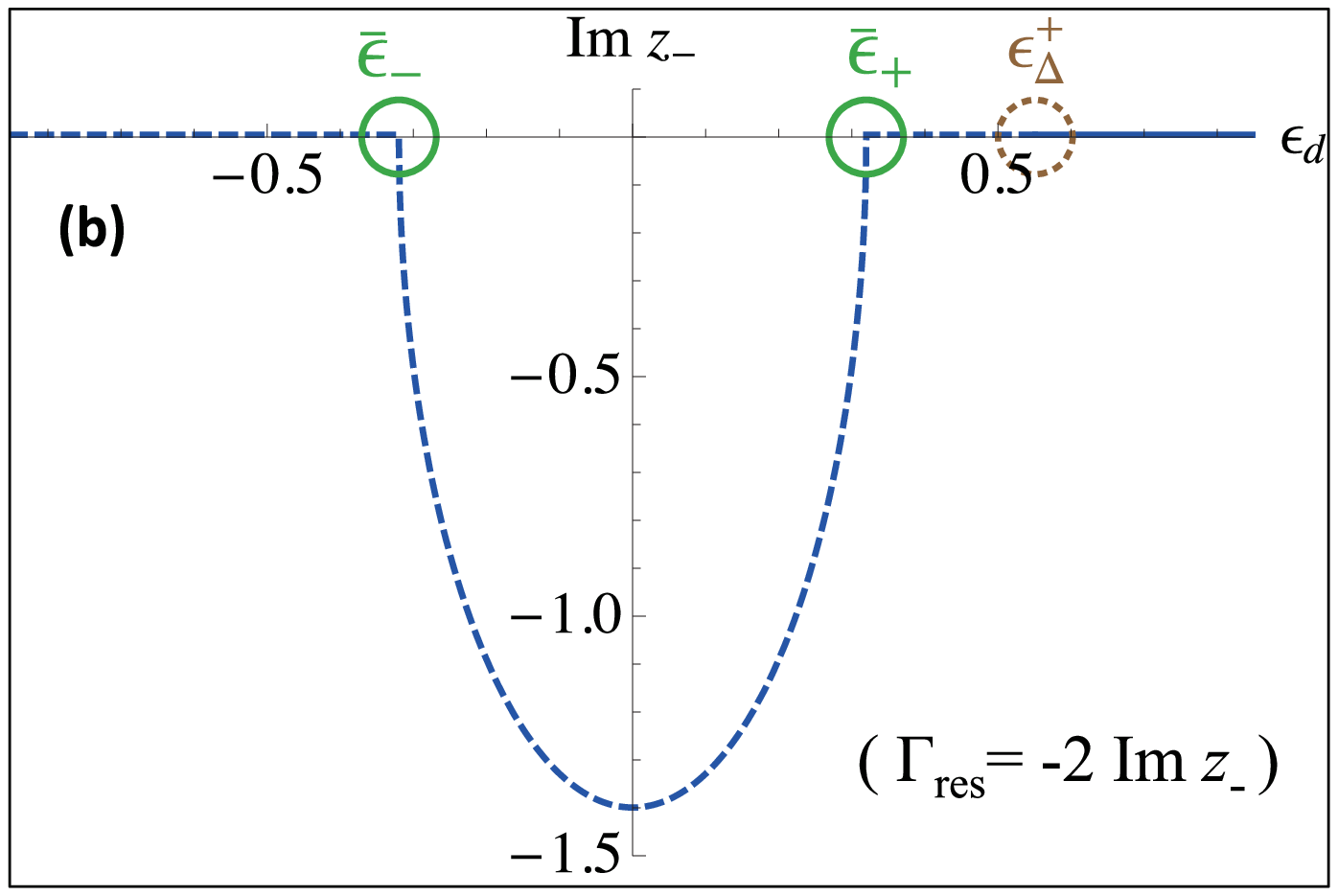}
 \includegraphics[width=0.45\textwidth]{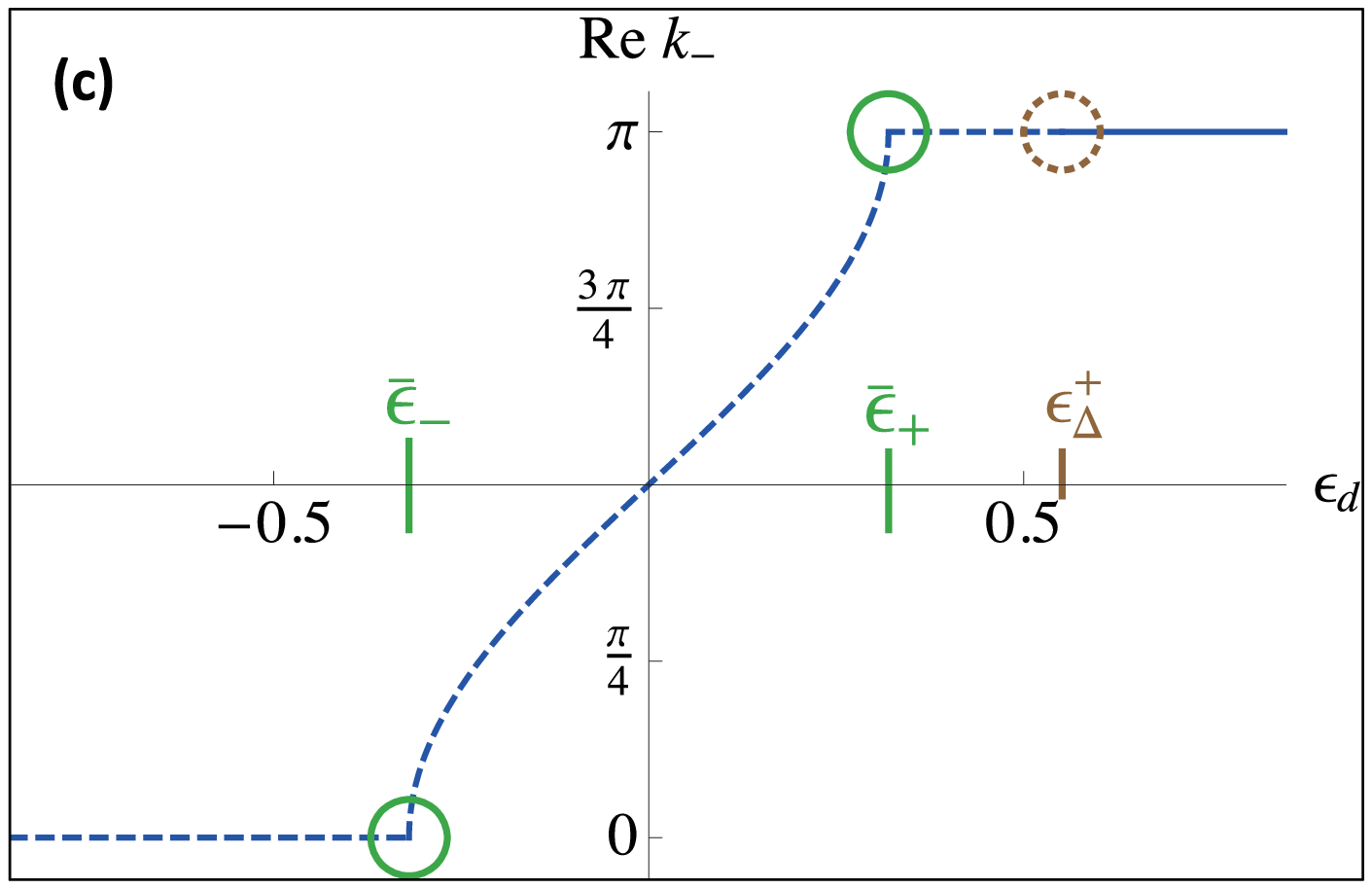}
 \includegraphics[width=0.45\textwidth]{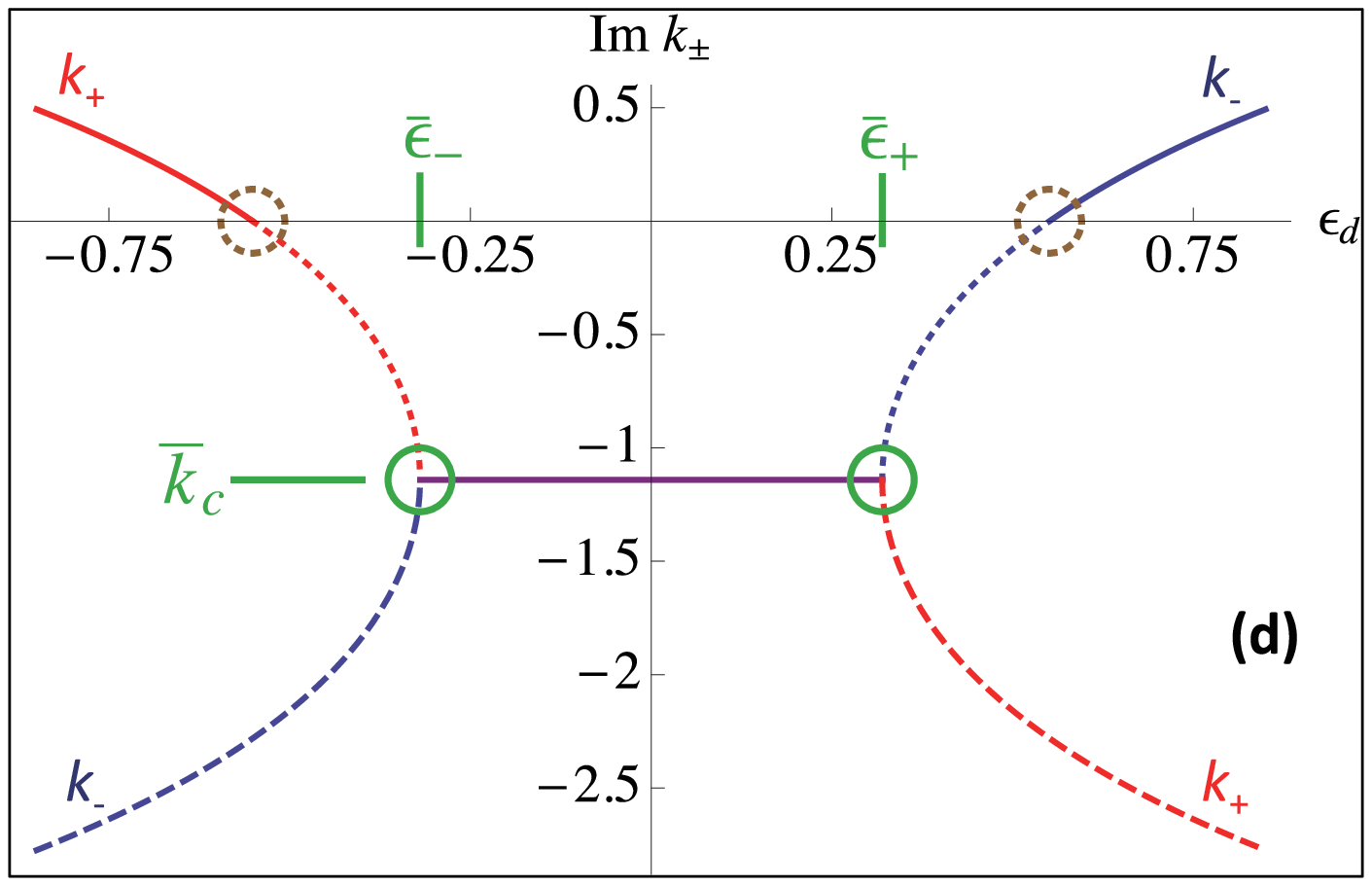}
\caption{ 
(Color online)
 (a) Real part of the $z_\pm$ spectrum (both eigenvalues), (b) imaginary part of $z_-$ (proportional to decay rate), (c) real part of the effective wave number $k_-$ for the $z_-$ eigenvalue, 
and (d) imaginary part of the effective wave numbers $k_\pm$ for both eigenvalues, 
plotted in each case as a function of $\epsd$ for fixed $g=0.67 < 1/\sqrt{2}$.  
}
 \label{fig:semi.spec}
\end{figure}

Before obtaining the fractional power series, we discuss the spectrum in the case $g < 1/\sqrt{2}$ and demonstrate that the resonant state emerges precisely at the EPs $\epspm$. 
The real and imaginary parts of the spectrum are plotted in Figs. \ref{fig:semi.spec}(a) and (b), respectively, as a function of $\epsd$ for fixed $g=.67$.  Solutions in the first (second) Riemann sheet are shown with a solid (dashed) curve.
In Fig. \ref{fig:semi.spec}(a) we find
for the case $\epsd > \epsilon_\Delta^+$ 
with
\begin{equation}
\epsilon_\Delta^\pm \equiv \pm (1 - g^2) ,
\label{semi.Edel}
\end{equation}
there exists exactly one bound state
given by the purely real solution $z_-  >  1$; this bound state occurs in the \emph{first} Riemann sheet in $z$ space.  For this case we find there also exists a real solution in the \emph{second} Riemann sheet that we call an anti-bound state; we discuss the properties of such a state momentarily.
If we decrease $\epsd$ to the value 
$\epsd = \epsilon_\Delta^+$ ($ > \epsp > 0$)
the bound state in the first sheet  turns around the branch point $z = 1$ (i.e., it brushes against the energy continuum)
before crossing into the second Riemann sheet as well \cite{Pastawski_footnote,Garmon_non-Mark,Economou}.
The crossover point is indicated in Fig. \ref{fig:semi.spec}(a) with a dotted brown circle.
For the negative side of the spectrum a similar process occurs in the vicinity of $\epsd = \epsilon_\Delta^-$ ($< \epsm < 0$), in which $z_-$ and $z_+$ switch roles.

This transition can be understood in terms of the evolution of the imaginary part of $k_-$, 
which from Eq. (\ref{semi.k.pm}) is easily shown to be positive-valued for $\epsd > \epsilon_\Delta^+$, as shown in Fig. \ref{fig:semi.spec}(d).  
This means that the wave function $\psi_- (x) = e^{ik_- x}$ is bounded for all $x$ (the electron tends to occupy levels close to the impurity).  If we decrease $\epsd$ the imaginary part of $k_-$ 
vanishes at $\epsd = \epsilon_\Delta^+$
and then becomes negative for $ \epsd < \epsilon_\Delta^+$, and hence the wave function $e^{ik_- x}$ has become unbounded.
Clearly, this type of solution no longer yields an ordinary bound state and hence we call it an 
\emph{anti-bound} state (or virtual state \cite{Pastawski_footnote}) 
in the second Riemann sheet in $z$ space; in this `state' the electron tends to occupy levels along the chain far away from the impurity.  The properties of the present model that induce this transition are detailed in Ref. \cite{Garmon_non-Mark}.

Physically, the transition from a bound to an anti-bound state is associated with an amplification of the long-time deviations from exponential decay \cite{Garmon_non-Mark} that are well known to occur in quantum systems \cite{Sudarshan,LFondaEtAl,Ch9_review}.  This transition is also associated with an amplification in the local density of states profile at the impurity site \cite{DBP08}.

We emphasize that $\epsilon_\Delta^\pm$ do not represent EPs since there is no coalescence of the wave functions here; in fact, these points are more closely related to the spectral singularities \cite{Longhi09,AM_PRL09} that may occur in the continuous spectrum of OQS if we relax our conditions on the Hamiltonian to allow for \emph{explicit} non-Hermiticity.  From a physical perspective, it has been shown that a diverging or vanishing reflection coefficient is obtained precisely at the point in parameter space at which the spectral singularity occurs.  These special points may be found in our prototype model by restricting the impurity energy $\epsd$ to complex values.  For a detailed analysis of the spectral singularities occurring in the present model, consult Ref. \cite{Longhi09}.

\begin{figure}
 \includegraphics[width=0.45\textwidth]{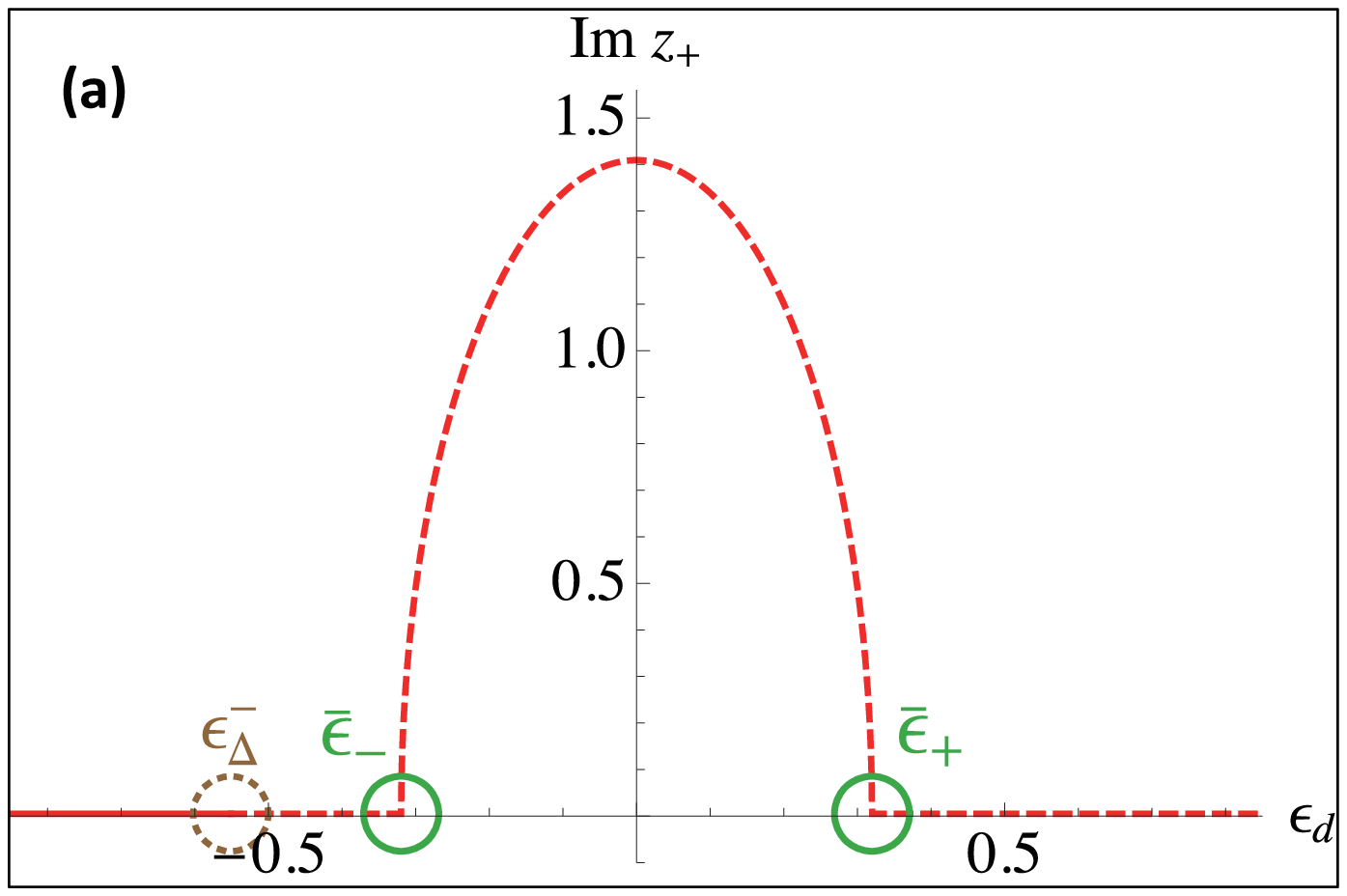}
 \includegraphics[width=0.45\textwidth]{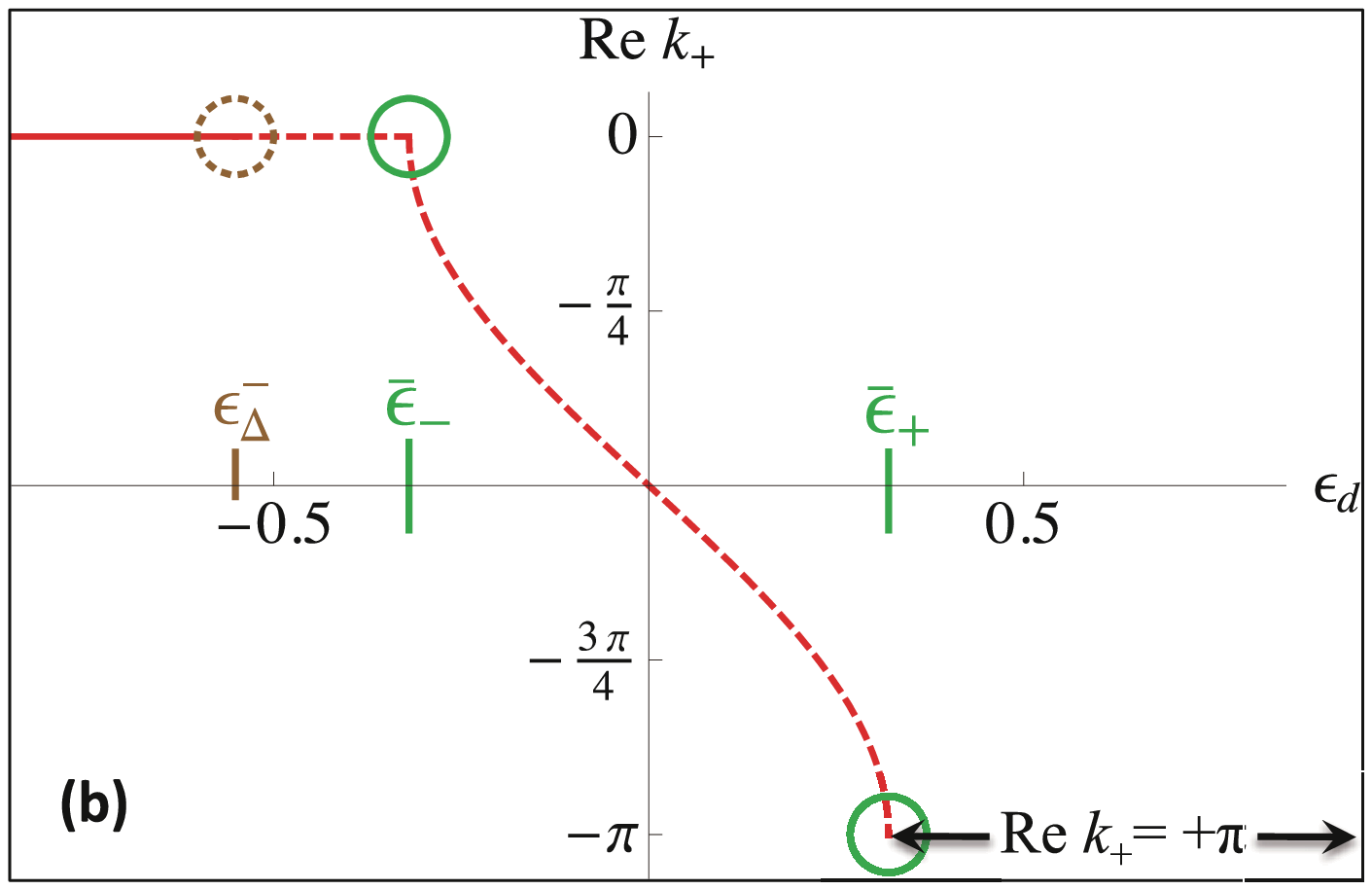}
\caption{ 
(Color online)
 (a) Imaginary part of $z_+$ and (b) real part of the effective wave number $k_+$, 
plotted as a function of $\epsd$ for fixed $g=0.67 < 1/\sqrt{2}$.  
}
 \label{fig:semi.spec.2}
\end{figure}

Now we discuss the emergence of the resonance state at the real-valued EPs $\epsd = \epspm$ (highlighted with solid green circles in Fig. \ref{fig:semi.spec}), which appears to be a general phenomena.
As shown above, in the region of the spectrum $\epsp < \epsd < \epsilon_\Delta^+$ 
there are two anti-bound states in the second Riemann sheet in $z$ space. 
At the precise values $\epsd = \epsp$ a different transition occurs as these two eigenvalues coalesce giving rise to a complex conjugate pair of solutions; this fact is apparent from the root appearing in the eigenvalue functions Eq. (\ref{semi.z.pm}).
The eigenvalue $z_-$ gives the resonant state with negative imaginary part (decay width), plotted in Fig. \ref{fig:semi.spec}(b).  Hence we find that the emergence of the resonance 
(associated with time irreversibility) 
occurs precisely at the EP.  
We finally point out in 
Fig. \ref{fig:semi.spec}(c) that for all $\epsd < \epsilon_\Delta^-$ ($\epsd > \epsilon_\Delta^+$) the real part of $k_-$ is $0$ ($\pi$);
for the anti-bound states, the wave vector always lies at one edge of the Brillouin zone.  It is only when the resonant state appears that $k_-$ takes a value inside the unperturbed channel of $k$-states, $k \in \left( 0 , \pi \right) $; from Eq. (\ref{semi.k.pm}) we find the exact value is given by 
$k_\textrm{res} = i \bar{k}_c +  \phi_\textrm{res}$ with $\bar{k}_c = \log \sqrt{1-2g^2}$ and
\begin{equation}
\phi_\textrm{res} = 
	\left\{ \begin{array}{lccc}
		\arctan \left( |\lambda^{1/2}| / \epsd \right) ,	&	&  	&  \epsm < \epsd < 0 ,	\\
			&	&	& 	\\
		 \arctan \left( - |\lambda^{1/2}| / \epsd \right) + \pi ,	&  	& 	&  0 < \epsd < \epsp ,
	\end{array}
	\right.   
\label{semi.phi.res}
\end{equation}
in which $\lambda \equiv \epsd^2 - (1 - 2g^2)$ (see Figs. \ref{fig:semi.spec}(c,d)).  Finally, for completeness we plot $\im \ z_+$ and $\re \ k_+$ in Figs. \ref{fig:semi.spec.2}(a,b), respectively.

We would now like to demonstrate the process of adiabatically encircling the EP.
Up until this point we have considered the chemical potential $\epsd$ at the impurity site to take on strictly real values.  However, let us  for a brief moment consider the possibility of a complex-valued 
$\epsd$, as has been studied for the present model in Ref. \cite{Longhi09}.  (We emphasize that in this reference the goal of the author is, unlike the present situation, to consider a complex chemical potential from the start, and then investigate the consequences of that generalization.)
We further imagine that the chemical potential at the impurity may be controlled by a gate potential, and hence we vary the impurity energy $\epsd$
around the EPs $\pm \sqrt{1 - 2 g^2}$ for a fixed $g$ value according to
$
\left( \epsd (\theta) \right)^2
	= 1 - 2g^2 + \delta e^{i \theta}   ,  \ \ \ 
$
with $\delta < 2 (1 - 2g^2)$ to avoid encircling the second EP as well.
Applying this parameterization to the eigenvalues Eq. (\ref{semi.z.pm}) we see that 
$\sqrt{\left( \epsd (\theta) \right)^2 - (1 - 2 g^2)} = \sqrt{\delta} e^{i \theta/2}$, which illustrates that $z_- \rightarrow z_+$ and vice versa as $\theta$ is varied from $0$ to 
$2 \pi$.  Hence the cycle structure Eq. (\ref{cycles}) is given by $\{z_-, z_+\}$ in the vicinity of either EP
for the present model.

In order to demonstrate the utility of our formalism and to motivate the succeeding QPT analogy we now obtain the fractional power eigenvalue expansion Eq. (\ref{kato.puiseux}) following a method from Ref. \cite{Knopp}.  
To do this we encode the variation of the system parameters about the EPs, in this case using
 $\zeta^2 = \lambda = \epsd^2 - (1 - 2g^2)$; we aim to perform expansions about the non-analytic 
points $\epsd = \epspm$, but the non-analyticity prevents us from doing so directly.
To overcome this problem, 
we write a new function for both eigenvalues $s = \pm$ at either EP as 
\begin{equation}
\phi_s (\epspm, \zeta) \equiv z_s (\epsd = \epspm, g) ,
\label{phi.s}
\end{equation}
in which we allow $g=\sqrt{\zeta^2 + 1-\epspm^2}/\sqrt{2}$ to vary as a function of $\zeta$; this effectively measures the distance from the EP in parameter space.
We can expand this new function $\phi_s$ in an ordinary Taylor series in $\zeta$
and then use the reverse parameter transformation to write the original eigenenergies as
\begin{equation}
z_s (\epsilon_\textrm{d}, g) 
	= 
		\bar{z}_\textrm{c}^\pm
		+ s \frac{1 - \epspm^2 }{2 \epspm^2 } \lambda^{1/2} 
			+ \frac{1}{2 \epspm} 
				\sum_{n=2}^\infty \left( \frac{s \lambda^{1/2}}{\epspm} \right)^n,
\label{semi.puiseux}
\end{equation}
with $\lambda = f_2 (\epsd, g)$ for comparison with Eq. (\ref{kato.puiseux}).  Keep in mind that each occurrence of $\epspm$ above specifies one of the two EPs, while $s = \pm$ refers to one of the two eigenvalues (i.e., Eq. (\ref{semi.puiseux}) is actually four equations).
This expression provides the realization of the generic expansion Eq. (\ref{kato.puiseux}) for the case of the semi-infinite chain with an endpoint impurity in which case $p$ takes the value $p=2$.


\section{QPT analogy for the appearance of the resonance at the EP}
\label{sec:qpt}

As a final comment on the EPs in OQS, we hypothesize a QPT analogy for the emergence of irreversibility at the EPs where the resonance appears.
This analogy is similar to that proposed in Ref. \cite{JMR99}, in which the decay rate $\Gamma = - 2 \ \im \ z_-$ is related to the order parameter.  We use the EP $\epsp$ as our example;
the analogy works out similarly for the negative side of the spectrum involving the EP $\epsm$.
Let us focus on the `non-critical case' $\epsp < \epsd < \epsilon_\Delta^+$, where $z_\pm$ are both real.  Here we have a pseudo-energy gap between the two anti-bound states given by 
$\Delta z \equiv z_+ - z_- \sim \lambda^{1/2} \sim (\epsd - \epsp)^{1/2}$ 
as shown in Eqs. (\ref{semi.z.pm}) or (\ref{semi.puiseux}); this pseudo-gap vanishes at the `critical point' $\epsd = \epsp$,
see Fig. \ref{fig:semi.spec}(a).  Then below the transition 
$\epsd < \epsp$ the decay width appears as $\Gamma_\textrm{res} \sim \lambda^{1/2} 
\sim (\epsd - \epsp)^{1/2}$, representing broken time-reversal symmetry.  We may carry the analogy further by defining a function that describes the correlations between the impurity site and an arbitrary chain site $x$ through the resonant state by
\begin{equation}
C_\textrm{res} (x ; d) \equiv
	\frac{1}{2 \pi i} \oint_{\mathcal C_-} dz \bra x \frac{1}{z - H} \ket d ,
\label{corr.fcn}
\end{equation}
where $\mathcal C_-$ represents a contour in the complex $z$ plane that surrounds only the  eigenvalue $z_-$ \cite{ZVW83}.
We evaluate this to find 
\begin{equation}
C_\textrm{res} (x ; d) 
	\sim e^{i k_\textrm{res} x} = e^{ \log \sqrt{1-2g^2} x} e^{i \phi_\textrm{res} x}.
\label{corr.fcn.eval}
\end{equation}

We recall that on the `non-critical' side $\epsd > \epsp$, $z_-$ is an anti-bound state with $\phi = \pi$, giving purely real-valued correlations.  
Meanwhile, on the `critical' side $0 < \epsd < \epsp$ we have 
$ \phi_\textrm{res} = \arctan \left(- |\lambda^{1/2}| / \epsd \right) + \pi$ from 
Eq. (\ref{semi.phi.res}), (plotted in Fig. \ref{fig:semi.spec}(c)), which gives complex-valued correlations.
Further, we see that $\phi_\textrm{res}$ takes a value within the continuum of $k$-states $\phi_\textrm{res} \in \left( 0, \pi \right)$ (the diagonalized states for the system if no impurity were present), signifying that the resonance state escapes the impurity and propagates along the chain with decay width $\Gamma_\textrm{res}$.

This is reminiscent of the dynamical phase transition in Ref. \cite{JMR99} that occurs when a resonance state, coupled to a large number of otherwise similar resonance states,
suddenly transitions from a narrow resonance to a wide resonance.
The order parameter associated with this process is proportional to the resonance decay width, which suffers a discontinuity in the first derivative at the transition point.  
Simultaneous with this transition, the wave function of this state ``collects all the corresponding components from the wave functions of all the other states.'' 

In the present case we instead have two anti-bound states (zero decay width) that spontaneously give rise at the EP to a resonance (finite decay width) and anti-resonance pair.
Meanwhile the correlations for the resonance appear similar to what we would find for the unperturbed $k$ states in the uncoupled model
\begin{equation}
C_0 (x ; k) = \bra x  \frac{1}{z - \HD - \HC} \ket k \sim e^{i k x},
\label{corr.fcn.x.k}
\end{equation}
with $k \in (0, \pi)$.
Hence the  resonance state in some sense mimics the $\ket k$ states, which are themselves nothing more than a collective oscillation of all the $| x \rangle$ states in the original site basis.
From Eq. (\ref{corr.fcn.eval}) we now introduce an inverse correlation length on the critical side of the transition as
$\xi^{-1} \equiv \phi_\textrm{res}$; further we note 
$\xi \sim \pi + \lambda^{-1/2} \sim \pi + (\epsd - \epsp)^{-1/2}$.
Since the decay width in this case also suffers a discontinuity in the first derivate at the EP, we interpret this as the order parameter varying as 
$\Gamma \sim \xi^{-z} \sim (\epsd - \epsp)^{1/2}$, which yields the dynamic exponent $z = 1$ \cite{Sachdev}; we propose that this exponent might be interpreted as representing the time dimension in which the symmetry is spontaneously broken with the appearance of exponential decay in association with the resonance state.

We note our argument above does not rest on the details of the prototype model but rather on generic features associated with the complex eigenvalue; for example
a similar transition occurs for a quite different physical situation in Ref. \cite{RSM07} and our QPT analogy should directly apply in the systems studied in Refs. \cite{PTG05,GNHP09,Longhi07,TGOP07,SHO11,DBP08}.  Indeed it seems that the resonant state in OQS should \emph{always} appear at a real-valued EP described by the expansion in Eq. (\ref{kato.puiseux}).

Recently, an experimental demonstration of precisely this type of transition has been performed in an electronic system consisting of a coupled \emph{LRC} circuit pair \cite{LCR_circuit11}.  In that system, one \emph{LRC} circuit is designed for amplification while the other provides a compensatory gain, resulting in a system with manifest parity-time $\mathcal{PT}$-symmetry \cite{BB98}.  By varying the system parameters, the authors of Ref. \cite{LCR_circuit11} are able to demonstrate a transition from real-valued eigenfrequencies (the ``exact $\mathcal{PT}$-symmetric phase'') to complex eigenfrequencies (the ``broken $\mathcal{PT}$-symmetric phase'').

It is worth noting at this point that although the Hamiltonian for our prototype model given in Eq. (\ref{ham.semi.site}) appears to be Hermitian, there is however an underlying non-Hermiticity associated with the continuum component of the system.  On a deeper level the system is in fact 
$\mathcal{PT}$-symmetric rather than Hermitian.  This underlying $\mathcal{PT}$-symmetric nature is revealed for the present system by writing the effective Hamiltonian $H_\textrm{eff}$ in Eq. (\ref{semi.H.eff}) (also see $H_\textrm{eff}$ re-written in terms of the $k$ variable in Eq. (\ref{semi.H.eff.gen})).  Indeed, applying the Hermitian transformation on Eq. (\ref{semi.H.eff.gen}) (with $F=1/2$ for the prototype model) reveals that the system is non-Hermitian \cite{Hatano_H_eff}, and indeed the non-Hermiticity comes from the lower-right entry for $H_\textrm{eff}$, which is the self-energy associated with the continuum.  However, $H_\textrm{eff}$ is indeed symmetric under the 
$\mathcal{PT}$ transformation.  This underlying  $\mathcal{PT}$-symmetric property is again a generic feature of OQS.

\section{Conclusion}
\label{sec:conc}

In this paper we have presented the basic mathematical properties of exceptional points occurring in the spectrum of open quantum systems, along with a technique for their location and analysis in these systems.  This technique includes our method for locating EPs in OQS detailed in 
Sec. \ref{sec:anal.tech}.  While the location method we have presented is oriented towards analytically solvable models, it nevertheless may prove useful in some numerical situations assuming that we may at least explicitly write the self-energy function $\Sigma (z)$ in integral form.  We have also outlined our method to obtain the fractional power expansion given in generic form in Eq. (\ref{kato.puiseux}), which itself has been generalized from the expression appearing in Kato's text Ref. \cite{Kato} (apparently this generalization is required to accommodate OQS, if perhaps not wider circumstances).

We have further applied these techniques to the case of a semi-infinite chain with an endpoint impurity.  We found that the resonant state eigenvalue emerges precisely at a real-valued EP, which should be a general result for OQS.  Further we offered a quantum phase transition analogy for the process in which one crosses this real-valued EP in parameter space.  According to this analogy the decay width, which experiences a discontinuity in its first derivative at the EP, plays the role of order parameter.  As the resonant state gives rise to time irreversibility via exponential decay, this appears to be related to the time-symmetry breaking mechanism in  $\mathcal{PT}$-symmetric systems, including OQS.


\appendix

\section{Derivation of the effective Hamiltonian $H_\textrm{eff}$ for the prototype model}
\label{app:H.eff}

In order to keep our presentation reasonably self-contained, we briefly outline the method from Ref. \cite{Hatano_H_eff} to obtain the effective Hamiltonian reported in Eq. (\ref{semi.H.eff}) from the full Hamiltonian for our prototype model given in Eq. (\ref{ham.semi.site}).  
We start by writing the wavefunction along the chain in generic form as
\begin{equation}
\psi (x) =
	C e^{i k x}  \ \ \ \ \ \ \ \     \mbox{for $x \ge 2$}
\label{semi.wave.fcn.C}
\end{equation}
with $C$ an undetermined normalization constant.  To solve the model from this point we find it useful to close the Schr\"odinger equation on the right with an arbitrary bra $\bra x $ to obtain
\begin{equation}
\bra x H \ket \psi
	= \epsilon \langle x | \psi \rangle
\label{x.generic.sch}
\end{equation}
in which $H$ is the full Hamiltonian from Eq. (\ref{ham.semi.site}), while $x$ might represent the impurity site or any site along the chain $x \in \{ \textrm{d}, 1, 2, \dots \}$.  
If we specify any site $x > 2$ and plug Eq. (\ref{ham.semi.site}) into Eq. (\ref{x.generic.sch}) then, after applying the usual commutation relations, we find 
\begin{equation}
- \frac{1}{2} \left( \ \langle x-1 | \psi \rangle + \langle x+1 | \psi \rangle \ \right)
	= \epsilon_k \langle x | \psi \rangle 
	.
\label{closed.x.sch.1}
\end{equation}
Then plugging in Eq. (\ref{semi.wave.fcn.C}) for the wave function we immediately find
$\epsilon_k = - \cos k$ for the dispersion along the chain.  Next we solve for the normalization constant 
$C$ by choosing $x = 2$  in Eq. (\ref{x.generic.sch}), which yields
\begin{equation}
- \frac{1}{2} \left( \ \langle 1 | \psi \rangle + C e^{3ik} \ \right)
	= \epsilon_k C e^{2ik}
	.
\label{closed.x.sch.2}
\end{equation}
After applying $\epsilon_k = - \cos k$ here we re-arrange to find $C = \langle 1 | \psi \rangle e^{-ik}$.

Finally, evaluating Eq. (\ref{x.generic.sch}) for the cases $x = 1$ and $x = \textrm{d}$ we obtain the matrix equation
\begin{equation}
\left( 
\begin{array}{cc}
		\epsilon_\textrm{d}	&  - \frac{g}{\sqrt{2}}	\\
		- \frac{g}{\sqrt{2}}	& - \frac{1}{2} e^{i k}
		\end{array}
	\right)
\left(  \begin{array}{c}
		\langle d | \psi \rangle  	\\
		\langle 1 | \psi \rangle
		\end{array}
	\right)
= z
\left(  \begin{array}{c}
		\langle d | \psi \rangle  	\\
		\langle 1 | \psi \rangle
		\end{array}
	\right) ,
\label{semi.H.eff.eqn}
\end{equation}
in which $\epsilon = z$ is the discrete eigenvalue associated with the impurity sector.
Applying the transformation $k = \cos^{-1} (-z)$ in the $2 \times 2$ matrix on the LHS of this equation we find
$- e^{i k} = z - \sqrt{z^2 - 1} = \Sigma (z)$, hence we obtain the effective Hamiltonian $H_\textrm{eff}$ reported in Eq. (\ref{semi.H.eff}) in the main text.  We now obtain the characteristic equation
by solving $\det \left( z - H_\textrm{eff} (z) \right) = 0$, as reported in Eq. (\ref{semi.disp}).

\section{Comment on the number of discrete solutions resulting from a general effective 
Hamiltonian $H_\textrm{eff}$}
\label{app:H.eff.n.solns}

In this Appendix we briefly comment on the number of solutions emerging from the characteristic equation under the effective Hamiltonian treatment (the reader should also consult App. D and H of 
\cite{SHO11} for similar discussions).
Let us consider a slightly more general form for  $H_\textrm{eff}$ of dimension two as
\begin{equation}
H_\textrm{eff} (z) 
	=
\left( 
\begin{array}{cc}
		\epsilon_\textrm{d}	&  - \frac{g}{\sqrt{2}}	\\
		- \frac{g}{\sqrt{2}}	& - F e^{i k}
		\end{array}
	\right).
\label{semi.H.eff.gen}
\end{equation}
(compare this with $H_\textrm{eff}$ from Eq. (\ref{semi.H.eff.eqn}) for the semi-infinite chain model with an endpoint impurity).
Note that we may just as well solve the characteristic equation in terms of $k$ rather than $z$ by writing
$\det \left( z(k) - H_\textrm{eff} (k) \right) = 0$, which gives
\begin{equation}
\left( e^{ik} + e^{-ik} + 2 \epsd \right) \left( (1 - 2 F) e^{ik} + e^{-ik} \right) - 2 g^2 = 0
\label{semi.disp.k}
\end{equation}
after applying the transformation $z(k) = - \cos k = (e^{ik} + e^{-ik})/2$ for a tight-binding chain.  Making the replacement $w = e^{ik}$ here and multiplying through by $w^2$ immediately yields a fourth order polynomial equation for $w$, hence a general effective Hamiltonian of dimension two should be expected to yield four discrete eigenvalues.  From this point it is not difficult to convince oneself that adding a single discrete impurity site increases the dimension of $H_\textrm{eff}$ by one and simultaneously introduces two new solutions to the characteristic equation.  Meanwhile, adding a new channel also increases the dimension of $H_\textrm{eff}$ by one but also requires us to solve a second polynomial equation, hence doubling the total number of solutions after the increase in the dimension of the effective Hamiltonian has already been taken into account.

However, our semi-infinite chain model with an endpoint impurity happens to have the special value $F=1/2$, which we note is the only value for $F$ under which the order of the polynomial Eq. (\ref{semi.disp.k}) is reduced from four to two.  Hence our present model represents a special case with regard to the number of discrete solutions (and therefore the number of EPs as well).

\begin{acknowledgements}
We would like to thank E. C. G. Sudarshan, G. Ordonez, J. Pixley and S. Kirchner for stimulating discussion on the QPT analogy and T. Petrosky for helpful comments on an earlier draft.  The research of S. G. was supported by CQIQC, the Sloan research fellowship of D. S. and MPI-PKS.  The research of N. H. was supported by Grant-in-Aid for Scientific Research No. 17340115 from the Ministry of Education, Culture, Sports, Science and Technology of Japan.
\end{acknowledgements}



\end{document}